\documentclass[apjl]{emulateapj}
\usepackage[sort&compress]{natbib}
\usepackage{hyperref}
\hypersetup{
    colorlinks=true,
    linkcolor=blue,
    citecolor=blue,
    filecolor=blue,
    urlcolor=blue
}
 \providecommand{\adsurl}[1]{\href{#1}{ADS}}

\usepackage{amsmath,amssymb}
\usepackage{mathptmx}
\usepackage{mathtools}
\usepackage[bottom]{footmisc}
\usepackage{graphicx}

\newcommand{\overbar}[1]{\mkern 1.5mu\overline{\mkern-1.5mu#1\mkern-1.5mu}\mkern 1.5mu}
\newcommand\ho{\ifmmode {\rm H{\small I}} \else H{\small I} \fi}
\newcommand\hh{\ifmmode {\rm H_2} \else H$_2$ \fi}
\def\msun{\ifmmode {\rm M_{\odot}}\else $\rm M_{\odot}$\fi}
\def\mpc{\ifmmode {\rm M_{\odot} \ pc^{-2}} \else $\rm M_{\odot} \ pc^{-2}$ \fi}
\def\tra{\ifmmode  \text{H{\small I}-to-H}_2\else H{\small I}-to-H$_2$ \fi}
\def\aG{\ifmmode {\alpha G}\else $\alpha G$ \fi}
\def\iuv{\ifmmode {I_{\rm UV}}\else $I_{\rm UV}$ \fi}

\newcommand\wt{\widetilde{w}}
\def\sg{\ifmmode \sigma_g \else $\sigma_g$ \fi}
\def\st{\ifmmode \widetilde{\sigma}_g \else $\widetilde{\sigma}_g$ \fi}
\newcommand\hd{\ifmmode \textrm{H{\small I}-dust} \else H{\small I}-dust \fi}

\begin{document}

\title{Analytic \tra Photodissociation Transition Profiles}
\author{
Shmuel Bialy$^\star$\altaffilmark{1} and Amiel Sternberg\altaffilmark{1}
}
\altaffiltext{1}
{Raymond and Beverly Sackler School of Physics \& Astronomy, Tel Aviv University, Ramat Aviv 69978, Israel}

\email{$^\star$shmuelbi@mail.tau.ac.il}

\slugcomment{Accepted for publication in the Astrophysical Journal}

\begin{abstract}
We present a simple analytic procedure for generating atomic (H{\small I}) to molecular $(\hh)$ density profiles for optically thick hydrogen gas clouds illuminated by far-ultraviolet radiation fields.
Our procedure is based on the analytic theory for the structure of one-dimensional H{\small I}/\hh photon-dominated regions, presented by Sternberg et al.~(2014).
Depth-dependent atomic and molecular density fractions may be computed for arbitrary gas density, far-ultraviolet field intensity, and the metallicity dependent H$_2$ formation rate coefficient, and dust absorption cross section in the Lyman-Werner photodissociation band.
We use our procedure to generate a set of \tra transition profiles for a wide range of conditions, 
from the weak- to strong-field limits, and from super-solar down to low metallicities.
We show that if presented as functions of  dust optical depth the \ho and \hh density profiles depend  primarily on the Sternberg ``$\aG$ parameter" (dimensionless) that determines the dust optical depth associated with the total photodissociated \ho column.
We derive a universal analytic formula for the \tra transition points as a function of just $\aG$. 
Our formula will be useful for interpreting emission-line observations of H{\small I}/\hh interfaces, for estimating star-formation thresholds, and for sub-grid components in hydrodynamics simulations.
\end{abstract}


\section{Introduction}
\label{sec: intro}
Stars form in dense molecular hydrogen (H$_2$) cores that are shielded from external  ultraviolet (UV) radiation. 
The H$_2$ plays a crucial role in the chemistry that occurs, and its formation leads to the subsequent production of other molecules such as CO, OH and H$_2$O \citep[e.g.,][]{Herbst1973, Sternberg1995, Tielens2013, vanDishoeck2013a, Bialy2015c}. 
These species are efficient coolants, and are able to cool the gas to very low temperatures $\sim 10 - 20$~K, 
reducing the Jeans masses and enabling fragmentation.
Star-formation may be triggered by the atomic-to-molecular $(\tra)$ phase transition. 
Alternatively, production of H$_2$ may be enhanced in the denser and optically thicker gravitationally collapsing components of the interstellar medium (ISM) \citep{Krumholz2011, Glover2012}.
In any case,  the \tra transition is a basic ingredient in any theory of star-formation and galaxy evolution.

Observations of molecular clouds in the Milky-way \citep[e.g.,][]{Allen2004, Barriault2010, Lee2015, Bialy2015b, Bihr2015}, as well as observations of external galaxies \citep[e.g.,][]{Wong2002, Blitz2004, Bigiel2008, Leroy2008, Schruba2011}, suggest that the atomic gas converts to molecular form once the \ho column density reaches a critical value.
Such a threshold is expected since the \ho is often a dissociation product of penetrating far-ultraviolet (FUV) radiation in a photon-dominated region (PDR). 
With increasing column density, the combination of H$_2$ self-shielding and dust absorption  attenuate the radiation field, and a conversion from \ho to H$_2$ occurs. 

Theoretically, the \tra conversion in PDRs has been widely investigated through analytic and numerical studies \citep[e.g.,~][]{Federman1979, VanDishoeck1986, Sternberg1988, Draine1996, Kaufman1999, Browning2003, McKee2010, Gnedin2014, Liszt2015},   
as well as in hydrodynamics simulations \citep[e.g.,~][]{Robertson2008, Gnedin2009, Glover2010, Bisbas2012, Dave2013, Thompson2014, Lagos2015}.

\citet[][hereafter S14]{Sternberg2014} presented analytic and detailed radiative transfer computations for the \tra transitions 
in one dimensional gas slabs irradiated by 
isotropic or beamed FUV radiation, and with emphasis on the build-up of the photodissociated \ho columns in optically thick gas.
They considered depth dependent multi-line H$_2$ photodissociation, and derived analytic formulae for the resulting total \ho columns. 
An important quantity is the dust opacity, $\tau_{\rm 1, tot}$, associated with the total \ho column. 
S14 refer to this as ``H{\small I}-dust" opacity and derived the analytic expression
\begin{equation*}
\tau_{1,{\rm tot}} \ \equiv \ \sg N_{1,{\rm tot}} \ = \ \ln \Big[ \frac{\aG}{2} \ + \ 1 \Big] \  ,
\end{equation*} for a slab exposed to beamed radiation.
Here $\sg$ (cm$^2$) is the metallicity dependent dust absorption cross-section per hydrogen nucleon  in the 11.2-13.6 eV Lyman-Werner (LW) photodissociation band, and $N_{\rm 1, tot}$ is the total \ho column produced by photodissociation in optically thick clouds, in which all of the radiation is absorbed.
The basic dimensionless parameter, $\aG$, that appears in this expression depends on several quantities (see Equation (\ref{eq: aG}) in \S~2) and is proportional to the ratio of the FUV intensity to gas density.
For realistic ISM conditions \aG may range from large ($\gg 1$) to small ($\ll 1$), and the \ho dust opacity may or may not be significant.

In this paper, we extend the S14 formalism to develop a simple analytic procedure for the construction of complete depth-dependent atomic and molecular density profiles for FUV illuminated gas.
The ``density profiles" are the local volume gas densities of \ho and \hh as functions of cloud depth, as parameterized by the total gas column density, or alternatively by the dust optical depth and/or visual extinction.
The analytic procedure we present in this paper provides a simple and quick method to generate such density profiles for a wide range of gas densities and radiation field intensities, and also the H$_2$ formation efficiencies and dust absorption cross-sections, as specified by the dust-to-gas mass ratio and the overall metallicity.
Following the formalism presented in S14, we show that the density profiles depend on the same two parameters, \sg and $\aG$, that determine the total \ho columns.
We use our procedure to generate \tra density profiles for a wide range of conditions, including for very low metallicities and dust-to-gas ratios.

Of particular interest are the gas columns and/or dust optical depths at which the conversion from atomic to molecular form occurs.
%
We will show that when expressed in terms of the dust optical depth, the transition points, $\tau_{\rm tran}$, depend almost entirely on just $\aG$, just as for the expression for $\tau_{\rm 1, tot}$.
Remarkably, this is irrespective of whether the conversion point is governed by \hh line self-shielding or \hd opacity.
We derive a universal fitting formula for the transition optical depth, given by
\begin{equation*}
\tau_{\rm tran} \ \equiv \sg N_{\rm tran} \ = \ \beta \ \ln \Big[ \Big(\frac{\aG}{2}\Big)^{1/\beta} + 1 \Big] \ .
\end{equation*} 
In this expression $N_{\rm tran}$ is the gas column, atomic {\it plus} molecular, at which the transition occurs.
For dust-to-gas ratios within 0.1 to 10 times the standard ISM value, we find that $\beta=0.7$ gives the transition point to high accuracy, for any $\aG$.

Our formula will be useful for interpreting observations since the H{\small I}/\hh transition layers are expected sources of ro-vibrational emissions from warm molecules \citep[e.g.,][]{Timmermann1996, Rosenthal2000, Rodriguez2001,  Allers2005, Shaw2009, Sheffer2011}. Our formula will also be useful as a ``sub-grid" ingredient in simulations that incorporate atomic to molecular conversion \citep[][]{Gnedin2009, MacLow2012, Dave2013, Valvidia2014, Thompson2014, Bahe2015}.
This is because with our formula, the transition point may be expressed in terms of external parameters only (i.e.~those that enter $\aG$) without requiring a solution for the detailed density profiles or the consideration of molecular self-shielding functions.

The structure of our paper is as follows. 
In \S \ref{sec: theory} we briefly review the S14 formalism, and elaborate on it to show that \aG and \sg fully determine the \ho and \hh density profiles. In \S \ref{sec: Profile constraction} we develop our analytic procedure for the construction of the atomic and molecular density profiles and we present formulae for the necessary auxiliary  functions.
In \S \ref{sec: profiles} we present our set of density profiles computed for a wide range of \aG and $\sg$, and we discuss their properties. 
We then derive our fitting formula for the atomic-to-molecular transition points, and also present attenuation factors.
In \S \ref{sec: star-formation} we obtain a threshold for the gas mass surface density required  for star-formation.
We summarize and conclude in \S \ref{sec: discussion}.

\section{Theory}
\label{sec: theory}
In this section we describe the basic analytic theory for the 
depth dependent variation of the atomic and molecular gas densities through steady state PDRs.
We follow the S14 theoretical framework for semi-infinite gas slabs.
S14 considered irradiation by isotropic or beamed fields. Here we assume beamed irradiation.

This theoretical analysis is for H$_2$ photodissociation by FUV Lyman-Werner photons. H$_2$ destruction by cosmic-ray and/or X-ray ionization requires a separate treatment \citep[e.g.,][]{Maloney1996, Meijerink2005, Bialy2015a}.

\subsection{Basic equations}
For steady state conditions, and for beamed radiation into one side of an optically thick slab,
\begin{equation}
\label{eq: H2 form_dest}
 R \ n \ n_1 \ =  \ \frac{1}{2} D_0 \ f_{\rm att} \ n_2 \ .
\end{equation}
The left-hand side is the H$_2$ formation rate per unit volume (cm$^{-3}$ s$^{-1}$), where $R$ is the H$_2$ formation rate coefficient (cm$^{3}$ s$^{-1}$).
The right-hand side is the local (attenuated) photodissociation rate per unit volume,
where $D_0$ is the free-space (unattenuated) photodissociation rate (s$^{-1}$), and $f_{\rm att}$ is the {\it depth dependent} attenuation factor accounting for H$_2$-self shielding and dust absorption.
In Equation (\ref{eq: H2 form_dest}), $n_1$ and $n_2$ are the local atomic and molecular volume densities (cm$^{-3}$) and $n \equiv n_1 + 2 n_2$  is the total gas density. 
For steady state this equality holds at every cloud depth.

The free space H$_2$ photodissociation rate is $D_0 = \sigma_d \overbar{F_{\nu}}$, where ${\overbar F_{\nu}}$ is the mean LW band flux density (photons cm$^{-2}$ s$^{-1}$ Hz$^{-1}$). As computed by S14, $\sigma_d =2.36\times 10^{-3}$~cm$^2$~Hz is the total H$_2$-line photodissociation cross-section, summed over all the lines. For a \citet{Draine1978} far-UV radiation spectrum ${\overbar F_{\nu}}=2.46 \times 10^{-8} I_{\rm UV}$~photon~cm$^{-2}$~s$^{-1}$~Hz$^{-1}$, and $D_0=5.8\times10^{-11} I_{\rm UV}$~s$^{-1}$, where $I_{\rm UV}$ is the radiation strength relative to the  Draine field, for which $I_{\rm UV}=1$. The factor 1/2 in Equation (\ref{eq: H2 form_dest}) accounts for removal of half of the free-space energy density by the optically thick slab.

Equation (\ref{eq: H2 form_dest}) is the fundamental relation for PDR theory, 
and it must be solved, either numerically or analytically, for the depth dependent atomic and molecular volume densities, and the conversion from \ho to H$_2$.

With increasing cloud depth the photodissociation rate is reduced by a combination of H$_2$-line self-shielding and dust opacity, and the molecular fraction increases.
In Equation (\ref{eq: H2 form_dest}), $f_{\rm att} \leq 1$ is the attenuation factor,
\begin{equation}
\label{eq: f_att}
f_{\rm att}(N_2, N) \ = \ f_{\rm shield}(N_2) \ \mathrm{e}^{-\tau} \ ,
\end{equation} and is the product of the dust attenuation term, $e^{-\tau}$, and the H$_2$-self shielding function $f_{\rm shield}(N_2)$. 
Here
\begin{equation}
\tau \ \equiv \ \sg \ N \ 
\label{eq: tau def}
\end{equation}
is the dust opacity in the LW band, 
where $N=N_1+2N_2$ is the hydrogen column density, in atoms plus molecules, 
 from the cloud surface to the given cloud depth, 
 and $\sg$ is the LW band dust absorption cross-section per hydrogen nucleon\footnote{In this paper we are following the S14 notation for which the subscript ``d" referes to photo-dissociation, and ``g" refers to dust-grains.
Thus, $\sigma_d$ is the photodissociation cross-section, and \sg is the dust-grain absorption cross section.}.
In Equation (\ref{eq: f_att}) we assume absorption (and pure forward scattering) by the grains, and neglect the (small) effects of back-scattering discussed by \citet[][see also S14]{Goicoechea2007}

  The visual extinction is related to $\tau$ through
 \begin{align}
 \label{eq: Av - tau}
 A_V \ &= \ 5.3 \times 10^{-22} \ N \ {\rm cm^2 \ mag} \nonumber\\
&= \ 0.28 \ \tau \ {\rm mag} \ .
\end{align} 
Both $\sigma_g$ and $A_V/N$ are proportional to the dust-grain surface area, and therefore depend strongly on the dust-to-gas ratio.
The dust-to-gas ratio may be further related to the overall gas metallicity.

For standard interstellar dust and assuming a linear relation between the dust-to-gas ratio and gas metallicity
\begin{equation}
\label{eq: sg Z}
\sg \ = \ 1.9 \times 10^{-21} \ \phi_g \ Z' \ {\rm cm^2} \ , 
\end{equation}
where $Z'$ is the metallicity relative to solar and $\phi_g$ is a factor of order unity  \citep[][S14]{Draine2003}.
For low metallicities the dust-to-gas ratio may scale superlinearly with the metallicity \citep{Remy-Ruyer2014} and $\sg$ will then scale accordingly. 

In our discussion below we consider \sg as a variable, and we define a normalized cross-section 
\begin{equation}
\label{eq: norm_cross_sec.}
\st \ \equiv \ \frac{\sg}{1.9 \times 10^{-21} \ {\rm cm^2}} \ ,
\end{equation} relative to the standard Galactic value.

As discussed in S14, $f_{\rm shield}$ depends on the molecular column, $N_2$, only, and its basic definition is
\begin{equation}
\label{eq: f_shield}
f_{\rm shield}(N_2) \ \equiv \ \frac{1}{\sigma_d} \ \frac{dW_{d}(N_2)}{dN_2} \ .
\end{equation}  
In this expression $W_d(N_2)$ is the bandwidth (Hz) of LW radiation absorbed in molecular photodissociations up to $N_2$, in a dust-free cloud, and is referred to as the ``{\it dust-free} H$_2$-line dissociation bandwidth".
Both $W_d(N_2)$ and its derivative $f_{\rm shield}(N_2)$ are 
very weakly dependent on the internal molecular excitation state (see S14) and hence may be computed in advance independent of external cloud parameters such as the gas density and field intensity.
We discuss analytic forms for $f_{\rm shield}(N_2)$ and $W_d(N_2)$ in \S \ref{sec: Profile constraction}.

In Equation (\ref{eq: H2 form_dest}) 
\begin{align}
\label{eq: alpha}
\alpha \ &\equiv \ \frac{D_0}{Rn} \ = \ \frac{\sigma_d \overbar{F_{\nu}}}{Rn}\nonumber\\
&= \ 5.8 \times 10^4 \ I_{\rm UV} \ \Big( \frac{10^{-17} \ {\rm cm^{3} \ s^{-1}}}{R} \Big) \Big( \frac{100 \ {\rm cm^{-3}}}{n} \Big)  \ ,
\end{align}
is the (dimensionless) ratio of the free-space H$_2$ photodissociation rate to the H$_2$ formation rate.
The parameter $\alpha$ is proportional to the ratio $I_{\rm UV}/n$, but it also depends on the H$_2$ rate coefficient $R$.
The H$_2$ formation rate coefficient depends on several quantities, including the gas-to-dust ratio and the gas and dust temperatures \citep[e.g.,][]{Hollenbach1979, Cazaux2004, LeBourlot2012}.
A characteristic value for $R$ is
\begin{equation}
\label{eq: R_H2}
R \ = \ 3 \times 10^{-17} \ \st \ {\rm cm^3 \ s^{-1}} \ ,
\end{equation}
with the further assumption that $R \propto \sg$ since both quantities depend on the dust-grain surface areas. This gives
\begin{equation}
\label{eq: alpha - R=dust}
\alpha \ = \ 1.9 \times 10^4 \ I_{\rm UV} \ \Big( \frac{100 \ {\rm cm^{-3}}}{n} \Big) \ \frac{1}{\st} \ .
\end{equation}
Therefore, as long as H$_2$ formation is dominated by dust catalysis, $\alpha$ is inversely proportional to the dust absorption cross-section.
For very low metallicities, H$_2$ formation is dominated by gas phase reactions \citep[e.g.,][]{Cazaux2004, Bialy2015a}, independent of $\sg$.

In terms of $\alpha$, Equation (\ref{eq: H2 form_dest}) may be rewritten as
\begin{equation}
\label{eq: n_1/n_2 alpha}
\frac{n_1}{n_2} \ = \ \frac{1}{2} \ \alpha \ f_{\rm att} \ .
\end{equation}
At the cloud boundary $f_{\rm att} \rightarrow 1$ and the \ho to \hh density ratio $n_1/n_2 \rightarrow 0.5 \alpha$. 
Since $n=n_1+2n_2$ the \ho and \hh fractions at the cloud boundary are $2n_2/n = 1/(1+0.25\alpha)$ and $n_1/n=0.25\alpha/(1+0.25\alpha)$. 
For most astrophysical conditions $\alpha \gg 1$ and the gas is predominantly atomic at the unshielded boundaries, with $n_1/n \simeq 1$ and  $2n_2/n \simeq 4/\alpha$.

The essential problem is to solve Equation (\ref{eq: n_1/n_2 alpha}) for $n_1$ and $n_2$ as functions of cloud depth, as parameterized by the gas column $N=N_1+2N_2$ (or visual extinction via Equation (\ref{eq: Av - tau})).
The solution is non-trivial because the attenuation factor depends on $N_2$ and $N$ in combination, not just on $N$ alone.
Thus, to obtain a solution for $n_1(N)$ and $n_2(N)$, one has to first solve for $N_2$ as a function of $N$.

\subsection{No dust absorption}
A simplifying case is the limit where dust absorption is negligible, i.e.~$\tau \ll 1$, for which $f_{\rm att} \rightarrow f_{\rm shield}$ and self-shielding dominates.
Assuming $Rn$ constant, and with Equation (\ref{eq: f_shield}) for $f_{\rm shield}$, 
and with $n_1/n_2=dN_1/dN_2$ for slab geometry, 
integration gives
\begin{equation}
\label{eq: N_1(N_2) no dust}
N_1(N_2) \ = \ \frac{1}{2} \ \alpha \ \frac{W_d(N_2)}{\sigma_d} \ = \frac{1}{2} \frac{\overbar{F_{\nu}} W_d(N_2)}{R n} \ ,
\end{equation} 
for the accumulated \ho column as a function of $N_2$.
For any $\alpha$, and given the precomputed $W_d(N_2)$, and with $N \equiv N_1+2N_2$, 
Equation (\ref{eq: N_1(N_2) no dust}) gives $N_1(N)$ and $N_2(N)$.

Equation (\ref{eq: n_1/n_2 alpha}) with $n \equiv n_1+2n_2$ then gives $n_1(N)$ and $n_2(N)$.
Thus, when dust absorption is negligible, the \ho and \hh density profiles are fully determined by the dimensionless parameter $\alpha$.
This solution is valid only up to gas columns $N \lesssim 5\times 10^{20}/\st$~cm$^{-2}$ (for which $\tau \lesssim 1$). For larger columns, dust absorption becomes significant.

\subsection{Inclusion of dust absorption}
With the inclusion of dust absorption, $f_{\rm att}$ becomes a function of both $N_2$ and $N$, and \sg enters as a parameter.
However, since $\mathrm{e}^{-\tau} \equiv \mathrm{e}^{-\sg N}=\mathrm{e}^{-\sg N_1} \times \mathrm{e}^{-2\sg N_2}$, 
 Equation (\ref{eq: n_1/n_2 alpha}) is separable and may be written as
\begin{align}
\label{eq: H2_form_dest_sep}
\mathrm{e}^{\sg N_1} \ dN_1 \ &= \ \frac{\alpha}{2}  \ f_{\rm shield}(N_2) \ \mathrm{e}^{-2 \sg N_2} \ dN_2 \  \nonumber\\
&=\ \frac{\alpha}{2} \ \frac{1}{\sigma_d} \ \frac{dW_g}{dN_2} \ dN_2 \ ,
\end{align}
where again $n_1/n_2=dN_1/dN_2$, as appropriate for slab geometry.
With this separation, and again assuming that $Rn$ is constant through the cloud,
 $N_1$ can still be expressed as a function of $N_2$, as follows.

The derivative on the right-hand side of Equation (\ref{eq: H2_form_dest_sep}) is of the function   
\begin{align}
W_g(N_2; \sg) \ &\equiv \ \int_0^{N_2} \frac{dW_d(N_2')}{dN_2'} \ \mathrm{e}^{-2\sg N_2'} \ dN_2' \nonumber\\
& = \ \sigma_d \ \int_0^{N_2} f_{\rm shield}(N_2') \ \mathrm{e}^{-2 \sg N_2'} \ dN_2' \ .
\end{align} 
Importantly, the exponential term in the integrand definition of $W_g(N_2; \sg)$ accounts for the dust opacity associated with the \hh column only (``H$_2$-dust"). 
Thus, as discussed in S14, $W_g(N_2; \sg)$ is the ``{\it H$_2$-dust limited} dissociation bandwidth".
$W_g(N_2; \sg)$ may be viewed as a curve-of-growth for the accumulating absorption of dissociating photons in a cloud that is fully molecular but also dusty.
The function $W_g(N_2; \sg)$ may be computed in advance for any assumed $\sg$, and is independent of the gas density or the radiation field strength.
S14 presented detailed numerical radiative transfer computations for $W_g(N_2; \sg)$.

For small \hh columns, H$_2$-dust absorption is negligible compared to H$_2$-line absorption so that $W_g(N_2) \rightarrow W_d(N_2)$, and the solutions approach the dust-free solutions given by Equation (\ref{eq: N_1(N_2) no dust}).
For larger \hh columns $W_g(N_2) < W_d(N_2)$ when some of the LW photons are absorbed by H$_2$-dust rather than in H$_2$ photodissociations.
For sufficiently large H$_2$ columns the dissociation bandwidth reaches a total (asymptotic) value.
Following S14 we refer to this asymptotic bandwidth as $W_{g,{\rm tot}}(\sg)$.
The asymptotic bandwidth is maximal for complete H$_2$ line overlap, and is reduced by H$_2$-dust opacity for sufficiently large $\st$.
The product $(1/2){\overbar F_{\nu}} W_{g,{\rm tot}}$ is then the {\it effective} dissociation photon flux absorbed by H$_2$ in photodissociations, excluding photons absorbed by H$_2$-dust, in fully molecular gas.

 It is convenient to also define the normalized (dimensionless) curve-of-growth 
 \begin{equation}
 \label{eq: Wtilde}
 \wt(N_2; \sg) \ \equiv \ \frac{W_g(N_2; \sg)}{W_{g,{\rm tot}}(\sg)} \ ,
 \end{equation} which has the property that for any $\sg$,  $\wt \rightarrow 1$ as $N_2 \rightarrow \infty$. 
In \S \ref{sec: Profile constraction} we plot the functions $W_g$, $\wt$, and $W_{g, {\rm tot}}$ (Figs.~\ref{fig: Wg} and \ref{fig: Wg_tot}) and also derive analytic forms for these quantities based on the numerical radiative transfer results presented by S14.

Equation (\ref{eq: H2_form_dest_sep}) may be integrated with $N_2$ the independent variable. This gives
\begin{equation}
\label{eq: N_1(N_2)}
N_1(N_2) \ = \ \frac{1}{\sg} \ \ln\Big[ \frac{\aG}{2} \ \wt(N_2; \sg) \ + \ 1 \Big] \ .
\end{equation} 
In this expression
\begin{align}
\label{eq: G}
G \ &\equiv \ \frac{\sg}{\sigma_d} \ W_{g,{\rm tot}} \nonumber\\
&= 3.0 \times 10^{-5} \ \st \ \Big( \frac{9.9}{1+8.9 \st} \Big)^{0.37} \ \  ,
\end{align} 
where we have used the analytic form we develop for $W_{g,{\rm tot}}(\st)$ in \S\ref{sub: Wgtot} below (Equation (\ref{eq: Wg_tot})).
As discussed by S14, the parameter $G$ is generally $\ll 1$, and may be viewed as the average H$_2$-self shielding attenuation factor, averaged over an H$_2$-dust optical depth $\sim 1$.
The product $D_0 G$ is then the characteristic shielded dissociation rate.

With the inclusion of dust absorption, Equation (\ref{eq: N_1(N_2)}) replaces Equation (\ref{eq: N_1(N_2) no dust}) for $N_1(N_2)$.
Together with $N=N_1+2N_2$ this gives $N_1$ and $N_2$ as functions of the gas column $N$.
This then gives the attenuation factor, $f_{\rm att}$, as a function of $N$ alone, determining the density profiles $n_1$ and $n_2$ versus the total gas column.

\subsection{Total \ho column and the $\aG$ parameter}
For $N_2 \rightarrow \infty$, $\wt \rightarrow 1$, and we recover the basic S14 expression for the total \ho column density
\begin{equation}
\label{eq: N_1_tot}
N_{1,{\rm tot}} \ = \ \frac{1}{\sg} \ \ln\Big[ \frac{\aG}{2} +1 \Big] \ .
\end{equation} 
This is for one side of an optically thick slab irradiated by a beamed field.
The \ho dust opacity associated with the total \ho column is then simply
\begin{equation}
\label{eq: tau_1_tot}
\tau_{\rm 1, tot} \ \equiv \ \sg N_{\rm 1, tot} \ = \ \ln\Big[ \frac{\aG}{2} \ + \ 1 \Big] \ . 
\end{equation}
In Equation (\ref{eq: N_1_tot}) and (\ref{eq: tau_1_tot}) 
the dimensionless parameter is
\begin{equation}
\label{eq: aG}
\aG \ \equiv \ \frac{D_0}{Rn} \ \frac{\sg}{\sigma_d} W_{g,{\rm tot}} \ = \ \frac{\sg \ {\overbar F}_{\nu} \ W_{g,{ \rm tot}}}{R n} \ ,
\end{equation} and it accounts for both H$_2$ self-shielding and dust absorption. 
This was a key result in S14 \citep[see also][]{Sternberg1988}.
The basic dimensionless parameter for $\tau_{\rm 1, tot}$ is always $\aG$, not $\alpha$ alone.
As shown by Equation (\ref{eq: aG}), \aG may be expressed as the ratio of the shielded dissociation rate to the \hh formation rate.
Alternatively, it is the ratio of the \hd absorption rate of the effective dissociation flux to the \hh formation rate.

 \begin{figure}[t]
	\centering
	\includegraphics[width=0.5\textwidth]{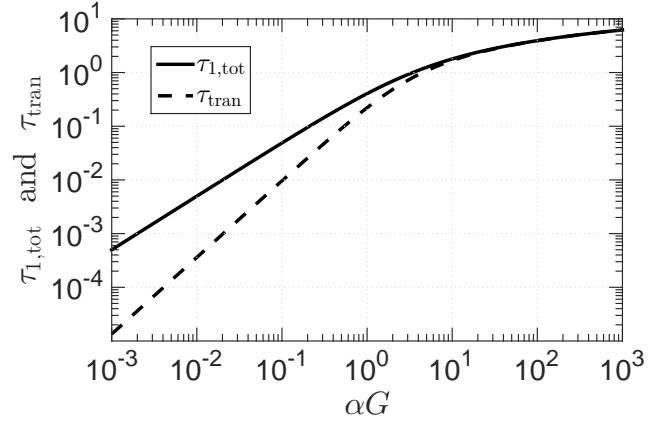}
	\caption{The total H{\tiny I}-dust opacity $\tau_{\rm 1,tot} \equiv \sg N_{\rm 1, tot}$ (solid) and the dust opacity at the \tra transition point $\tau_{\rm tran} \equiv \sg N_{\rm tran}$ (dashed), as functions of the basic dimensionless parameter $\aG$ (see text).
	}
	\label{fig: N1_tot_tau1_tot}
\end{figure}

Using Equations (\ref{eq: alpha}) and (\ref{eq: G}) for $\alpha$ and $G$ we may write \aG in the normalized form
\begin{align}
\label{eq: aG param with R}
\alpha G = &1.76  \ I_{\rm UV} \ \st \  \Big( \frac{10^{-17} \ {\rm cm^3 \ s^{-1}} }{R} \Big) \Big( \frac{100 \ {\rm cm^{-3}} }{n} \Big)\nonumber\\
& \times \Big( \frac{9.9}{1+8.9 \st} \Big)^{0.37}  \ .
\end{align}
For H$_2$ formation on dust grains (Equation (\ref{eq: R_H2})),
\begin{equation}
\label{eq: aG param}
\aG \ = \ 0.59 \ I_{\rm UV} \Big(\frac{100 \ {\rm cm^{-3}}}{n} \Big) \  \Big( \frac{9.9}{1+8.9 \st} \Big)^{0.37}  \ .
\end{equation} 
Like $\alpha$, the product $\aG$ is also proportional to the ratio $I_{\rm UV}/n$, but the prefactor is of order unity.
Both large and small \aG are relevant for the realistic range of interstellar conditions.
The remaining factor $1/(1+8.9\st)^{0.37}$ accounts for the reduction in the total dissociation bandwidth by H$_2$-dust absorption, which becomes important for $\st \gtrsim 0.1$.
Because \aG itself depends on $\st$, the ratio $I_{\rm UV}/n$ must be adjusted if \aG is held constant for varying $\st$.
For example, for $\aG=1$, $I_{\rm UV}/n= 9\times 10^{-3}$, $2 \times 10^{-2}$, and $4\times 10^{-2}$~cm$^3$, for $\st = 0.1$, 1, and 10.

In Fig.~\ref{fig: N1_tot_tau1_tot} we plot $\tau_{1,{\rm tot}}$ as a function of $\aG$ (solid curve).
For $\aG \ll 1$, $\tau_{\rm 1, tot}$ increases linearly with $\aG$. This is the ``weak-field" limit.
For $\aG \gtrsim 1$, $\tau_{\rm 1, tot} \gtrsim 1$, \hd dominates the absorption of the LW-band radiation, and the \ho column is ``self-limited". 
This is the ``strong-field" limit.

For $\aG \ll 1$, $\tau_{\rm 1, tot} \ll 1$ and \hd absorption is negligible. In this limit $N_1(N_2) \rightarrow  0.5\aG \wt/\sg = 0.5 \alpha W_g/\sigma_d$ (notice that \sg drops out).
If H$_2$-dust is also negligible $W_g \rightarrow W_d$ and we recover Equation (\ref{eq: N_1(N_2) no dust}) for the dust-free conditions.

In Fig.~\ref{fig: N1_tot_tau1_tot} we also plot (dashed curve) the analytic expression we derive in \S\ref{sub: transition points} for the dust optical depth $\tau_{\rm tran}$, at which the \tra transition occurs (see Equation (\ref{eq: tau_g_tran approx})).
We first present our procedure for generating the atomic and molecular density profiles.
These will determine the transition points as functions of $\aG$ and $\st$.

\subsection{Time-scales}
The above analysis, beginning with the formation-destruction Equation (\ref{eq: H2 form_dest}) is for steady state conditions. The time-scale to reach steady-state is given by
\begin{equation}
\label{eq: time scale}
t_{\rm eq} \ = \ \frac{1}{D \ + \ 2 \ R \ n} \ ,
\end{equation} 
where $D$ is the local (attenuated) photodissociation rate and $Rn$ is the formation rate.
When $D/(2Rn) \gg 1$, $t_{\rm eq}$ is the dissociation time $\simeq 1/D$, and the gas becomes atomic on a short time-scale. 
For example, in free-space $D=D_0$ and $t_{\rm eq} \approx 5.5 \times 10^2/I_{\rm UV}$~yr.
When $D/(2 R n) \ll 1$, $t_{\rm eq}$ is the molecular formation time $\simeq 1/(2 R n) \approx 5\times10^8/n$~yr.
Because the molecular formation time is long, non-equilibrium effects \citep[e.g.,][]{Liszt2007}
may become important beyond the \tra transition points.

\section{Analytic procedure for generating profiles}
\label{sec: Profile constraction}
Differentiating Equation (\ref{eq: N_1(N_2)}) gives a formal expression for the atomic-to-molecular density ratio as a function of just the {\it molecular} column,
\begin{equation}
\label{eq: profiles}
\frac{n_1}{n_2}(N_2)  \ =  \  \frac{1}{\sg} \ \times \ \frac{\aG \ \wt~'(N_2;\sg)}{\aG \ \wt (N_2; \sg)+2} \ ,
\end{equation} where $\wt~' \equiv d\wt/dN_2$.
Together with Equation (\ref{eq: N_1(N_2)}) this can be converted to $n_1/n_2$ as a function of the {\it total} gas column $N$, since $N=N_1(N_2)+2N_2$.
Because the functions on the right-hand-sides of Equations (\ref{eq: N_1(N_2)}) and (\ref{eq: profiles}) depend on just \sg and $\aG$, so do $n_1/n_2$, $n_1/n$ and $n_2/n$, all as functions of $N$.

Equation (\ref{eq: profiles}) can be reexpressed in the computationally convenient form
\begin{align}
\label{eq: profiles2}
\frac{n_1}{n_2}(N_2)  \ &=  \  \frac{\alpha \ f_{\rm shield}(N_2) \ \mathrm{e}^{-2\sg N_2}}{\aG \ \wt(N_2; \sg) + 2} \nonumber\\
&= \ \Big(\frac{\sigma_d}{\sg}\Big) \ \frac{ \aG  \ f_{\rm shield}(N_2) \ \mathrm{e}^{-2  \sg N_2}}{ \aG \ W_g(N_2; \sg) \ + \ 2W_{g,{\rm tot}}(\sg)}  \ .
\end{align} 
The depth-dependent atomic and molecular densities may then be constructed in a simple procedure:
\begin{enumerate}
\item Select values for \sg and $\aG$ (Equations (\ref{eq: norm_cross_sec.}) and (\ref{eq: aG param})).
\item Evaluate $n_1/n_2$ as a function of $N_2$ using Equation (\ref{eq: profiles2}), with the analytic forms for $f_{\rm shield}$, $W_g$, and $W_{g,{\rm tot}}$ (Equations (\ref{eq: f_shield DB96}), (\ref{eq: fit}), and (\ref{eq: Wg_tot}) below).
\item Compute $N_2(N)$ using Equation (\ref{eq: N_1(N_2)}) for $N_1(N_2)$ together with the relation $N=N_1+2N_2$.
\item Convert $n_1/n_2$ versus $N_2$ to $n_1/n_2$ as a function of $N$.
\item Obtain $n_1/n$ and $2n_2/n$ as functions of $N$, assuming constant gas density $n \equiv n_1+2n_2$. 
\end{enumerate}
In the reminder of this section we provide the analytic expressions for $f_{\rm shield}$, $W_g$, and $W_{g,{\rm tot}}$, required to evaluate Equation (\ref{eq: profiles2}).

\subsection{Self-shielding function $f_{\rm shield}(N_2)$}
\label{sub: fitting for f}
The self-shielding function $f_{\rm shield}(N_2)$ as defined by Equation (\ref{eq: f_shield}) was computed via multi-line radiative transfer in S14.
The results (see Fig.~2 of S14) are in excellent agreement with the \citet{Draine1996} fitting function
\begin{align}
\label{eq: f_shield DB96}
f_{\rm shield} \ = \ &\frac{0.965}{(1+x/b_5)^2} \ + \ \frac{0.035}{(1+x)^{0.5}}  \nonumber\\
&\times ֿ\exp[-8.5\times 10^{-4} (1+x)^{0.5} ]\ .
\end{align} Here, $x\equiv N_2/(5\times10^{14} \ {\rm cm^{-2}})$ and $b_5 \equiv b/(10^5 \ {\rm cm \ s^{-1}})$ is the normalized Doppler parameter.
In our computations we assume $b_5=2$ as a compromise between purely thermal and turbulent broadened linewidths. As we discuss further below (\S \ref{sub: profiles vs tau}) the positions of the transition points depend weakly on $b$.

\begin{figure}[t]
	\centering
	\includegraphics[width=0.5\textwidth]{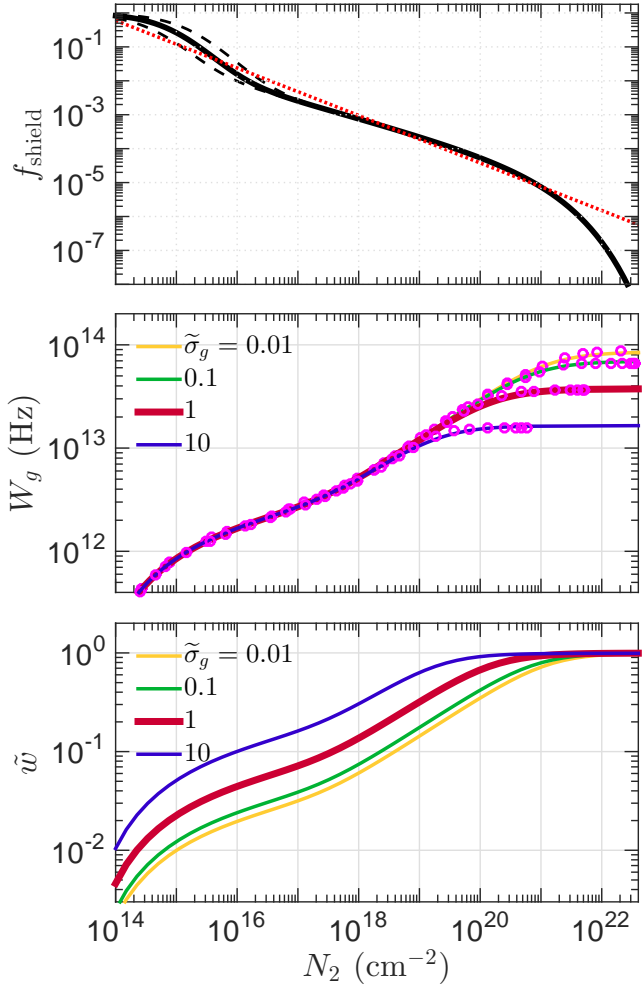}
	\caption{Top: The H$_2$-self shielding function $f_{\rm shield}$ as a function of the H$_2$ column $N_2$, for three values of Doppler parameters $b_5=2$ (solid) $b_5=1$ (upper dashed), and $b_5=4$ (lower dashed). The red dotted line is the power law approximation $f_{\rm shield} \propto N_2^{-0.7}$.
	Middle:
	The H$_2$-dust limited dissociation bandwidth $W_g$, for various values of the normalized dust cross section $\st$. The circles are the radiative transfer calculations and the solid colored curves are our fit, given by Equation (\ref{eq: fit}). 
	For small $N_2$, $W_g$ is independent of \st and is similar to a single-line absorption.
	For large $N_2$, all the UV photons are absorbed and $W_g$ reaches an asymptotic value $W_{g,{\rm tot}}$.
	Bottom: the normalized curve of growth $\wt \equiv W_g/W_{g,{\rm tot}}$.	}
	\label{fig: Wg}
\end{figure}

In Fig.~\ref{fig: Wg} (upper panel) we plot $f_{\rm shield}$ as a function of $N_2$ for $b_5=1, 2$, and 4.
The onset of self-shielding occurs at small H$_2$ columns, $\sim 10^{14}$~cm$^{-2}$, the exact value depending on $b$.
At $N_2 \gtrsim 10^{17}$~cm$^{-2}$ the absorption occurs out of the radiative damping wings and the self-shielding is independent of the Doppler parameter. 
Finally, at very large columns, $N_2 \gtrsim 10^{22}$~cm$^{-2}$, the absorption lines overlap and $f_{\rm shield}$ drops sharply. 

As we discuss in \S\ref{sub: transition points} below in our analysis of the transition points, a simple power-law approximation $f_{\rm shield} \propto N_2^{-\beta},$ is useful.
For the range $10^{14} < N < 10^{21.5}$~cm$^{-2}$, a power law index $\beta = 0.7$ provides a good approximation. 
This is shown as the red dotted line in Fig.~\ref{fig: Wg} (upper panel).

\subsection{H$_2$-dust limited dissociation bandwidth $W_g(N_2; \sg)$}
\label{sub: Wg}
S14 presented radiative transfer computations for the H$_2$-dust limited dissociation bandwidth, $W_g(N_2; \sg)$, 
for four discrete values of $\sg$, ranging from 
$\st = 0.01$ to 10.
Recall, $\st$ is the normalized dust cross section, normalized to $1.9 \times 10^{-21} \ {\rm cm^2}$.
Those computations were carried out using the Meudon PDR code\footnote{Publicly available at \href{http://pdr.obspm.fr}{http://pdr.obspm.fr}}.
 The numerical S14 results for $W_g(N_2; \sg)$ are the circles in Fig.~\ref{fig: Wg} (middle panel). For our purposes we require analytic expressions that vary smoothly with $N_2$ for {\it any} \sg within this range. We find that an excellent fit to the S14 calculations is
\begin{equation}
\label{eq: fit}
W_g(N_2; \st) \ = \ a_1 \ln \Big[ \frac{a_2 + y}{1 + y/a_3} \Big] \ \Big( \frac{1+y/a_3}{1+y/a_4} \Big)^{0.4} \ ,
\end{equation}
where 
\begin{align*}
y \ \ &\equiv \ \frac{N_2}{10^{14} \ {\rm cm^{-2}}} \nonumber\\ 
a_1 \  &= \  3.6 \times 10^{11} \ {\rm Hz} \nonumber\\
a_2 \  &= \  0.62 \nonumber\\
a_3 \  &= \  2.6 \times 10^{3}  \nonumber\\
a_4 \  &= \  1.4 \times 10^7  \ \times \ (1+ 8.9 \st)^{-0.93} \ .
\end{align*} 
The analytic fits are displayed as the solid curves in Fig.~\ref{fig: Wg} (middle panel).
Our analytic formula for $W_g(N_2; \sg)$ is accurate to within 6 \% of the numerical results for $N_2 > 10^{14} \ {\rm cm^{-2}}$ and for any $\st \leq 10 $.

The form of our fitting function Equation (\ref{eq: fit}), is physically motivated by ``the curve of growth" behavior of the dissociation bandwidth, which up to the H$_2$-dust cutoff closely resembles that for a simple absorption line.
For small H$_2$ columns ($y \lesssim a_2$), the photons are absorbed in optically thin Doppler-cores and $W_g \propto N_2$, i.e~the linear regime.
For intermediate columns ($a_2 \lesssim y \lesssim a_3$), $W_g \propto \ln[N_2]$, corresponding to the flat part of the curve-of-growth. At large columns ($a_3 \lesssim y \lesssim a_4$) absorption is dominated by the damping wings, and $W_g$ scales as a power law of $N_2$. Finally, for very large columns ($y \gtrsim a_4 $), the LW photons are either fully absorbed in overlapping H$_2$ lines or by H$_2$-dust. Either way, $W_g$ flattens. 
H$_2$-dust dominates the cutoff for $\st \gtrsim 0.1$ and $a_4$ is then sensitive to $\st$.
For $\st \ll 0.1$, H$_2$-dust is negligible and the entire LW band is absorbed in overlapping H$_2$ lines. In this limit $W_g(N_2; \sg) \rightarrow W_d(N_2)$ and is maximal, reaching $8.8 \times 10^{13}$~Hz, independent of $\sg$.

	\begin{figure}[t]
	\centering
	\includegraphics[width=0.5\textwidth]{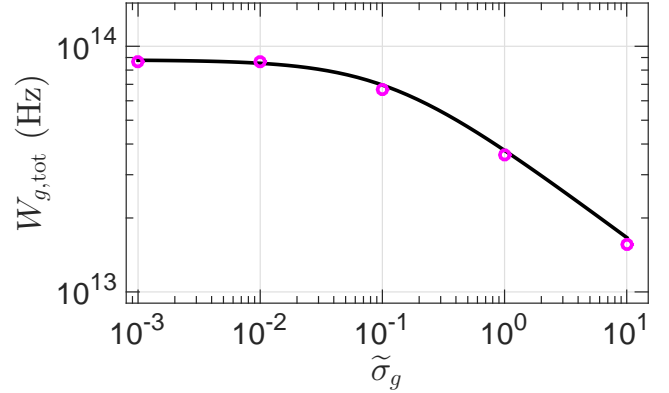}
	\caption{The total H$_2$-dissociation bandwidth $W_{g,{\rm tot}}$ as function of the normalized dust cross section $\st$. The magenta circles are the explicit radiative transfer calculations of S14, and the solid curve is our analytic formula, Equation (\ref{eq: Wg_tot}).
	}	
	\label{fig: Wg_tot}
\end{figure}

\subsection{Total dissociation bandwidth $W_{g,{\rm tot}}(\sg)$}
\label{sub: Wgtot}

As $N_2$ becomes large $W_g(N_2; \sg)$ approaches the total H$_2$-dust limited dissociation bandwidth $W_{g,{\rm tot}}(\sg)$. 
The circles in Fig.~\ref{fig: Wg_tot} are the radiative transfer numerical results as computed by S14 for $W_{g,{\rm tot}}(\sg)$. S14 also presented the fitting function 
\begin{equation}
\label{eq: Wg_tot_S14}
W_{g,{\rm tot}}^{\rm S14}(\st) \ = \ \frac{9.9\times10^{13} \ {\rm Hz}}{1+(2.6\st)^{1/2}} \ .
\end{equation}
Here we adopt an alternative and more accurate expression
\begin{equation}
\label{eq: Wg_tot}
 W_{g, {\rm tot}}(\st) \ = \ \frac{8.8 \times 10^{13} \ {\rm Hz}}{(1+8.9\st)^{0.37}}  \ ,
\end{equation}
plotted as the solid curve in Fig.~\ref{fig: Wg_tot}.
This expression is obtained by taking the limit $N_2 \rightarrow \infty$ in our formula for $W_g(N_2; \st)$ (Equation (\ref{eq: fit})).
For large $\st$, $W_{g,{\rm tot}}$ decreases as
an increasing fraction of the LW band radiation is absorbed by H$_2$-dust, before complete line overlap occurs. 
For $\st \ll 0.1$, H$_2$-line absorption dominates, and $W_{g,{\rm tot}} \rightarrow  W_{d,{\rm tot}} \equiv 8.8 \times 10^{13}$~Hz, independent of $\st$.

Given our analytic forms for $W_g$ and $W_{g,{\rm tot}}$ we can compute the normalized curve of growth $\wt(N_2) \equiv W_g(N_2)/W_{g,{\rm tot}}$.
In the lower panel of Fig.~\ref{fig: Wg} we plot $\wt$ for $\st=0.01, 0.1, 1$ and 10.
By definition $\wt$ is normalized such that it approaches unity for large $N_2$.
Therefore, for any $N_2$ prior to saturation, $\wt$ is larger for larger values of $\st$.

%

\section{Density profiles and transition points}
\label{sec: profiles}
\begin{figure*}
	\centering
	\includegraphics[width=1\textwidth]{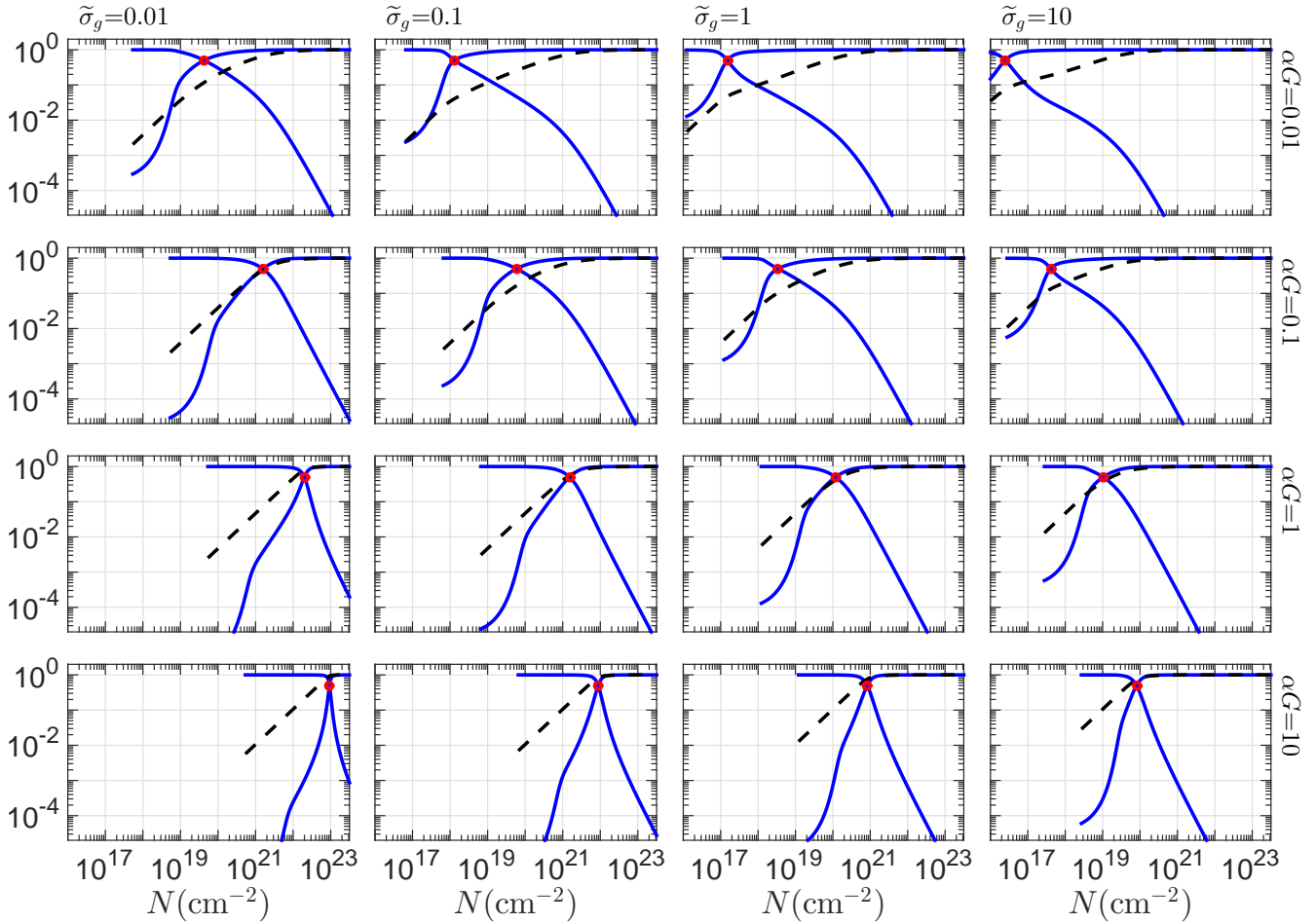}
	\caption{The H{\tiny I} and \hh density profiles $n_1/n$ and $2n_2/n$ (blue solid curves) and the normalized accumulated H{\tiny I} column $\widetilde{N_1}(N)$ (dashed black curves), as functions of the total hydrogen column $N$, for the $4\times 4$ combinations of the basic parameters $\aG = 0.01$, 0.1, 1, 10, and $\st=0.01$, 0.1, 1, 10.
	The profiles shown are evaluated through the analytic procedure presented in \S \ref{sec: Profile constraction}.
	The H{\tiny I}-to-H$_2$ transition (red points) occurs at larger depths with increasing $\aG$ (stronger UV fields), or decreasing $\st$ (weaker absorption). The H{\tiny I}-to-H$_2$ transition is gradual for $\aG \ll 1$ and becomes sharp for $\aG \gtrsim 1$.
	}
	\label{fig: profiles_N}
\end{figure*}
 \begin{figure*}
	\centering
	\includegraphics[width=1\textwidth]{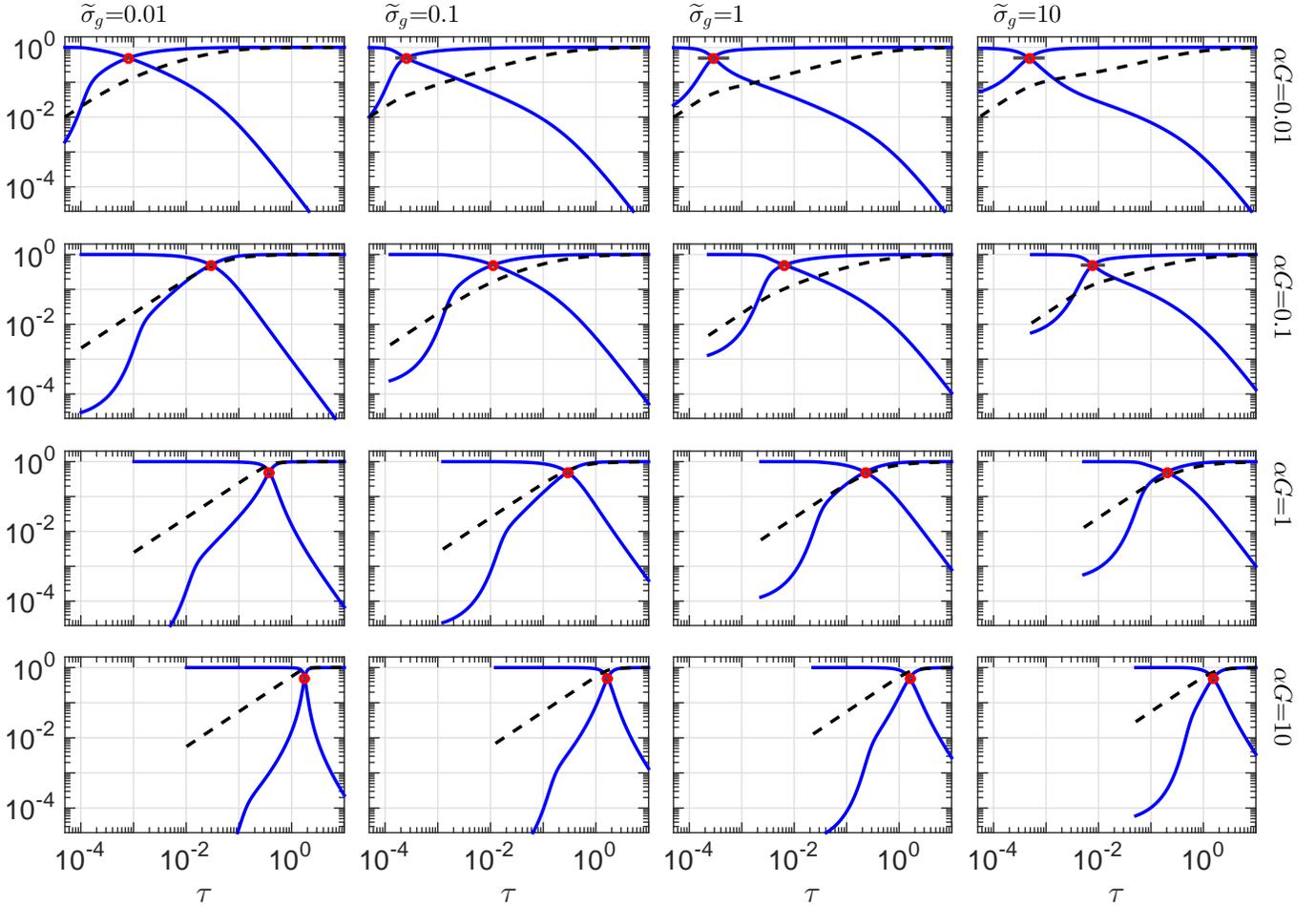}
	\caption{The H{\tiny I} and \hh density profiles $n_1/n$ and $2n_2/n$ (blue solid curves) and the normalized accumulated H{\tiny I} column $\widetilde{N_1}$ (dashed black curves), as functions of the dust opacity $\tau \equiv \sg N$, for the $4\times 4$ combinations of the basic parameters $\aG = 0.01$, 0.1, 1, 10, and $\st=0.01$, 0.1, 1, 10.
The horizontal bars (mainly seen in the upper right panels) indicate the variation of the transition points for varying Doppler parameter $b_5=1$ to 4.  
The profile shapes (sharp versus gradual) and the positions of the H{\tiny I}-to-H$_2$ transition points are determined mainly by the $\aG$ parameter.
}
	\label{fig: profiles_tau_log}
\end{figure*}
 \begin{figure*}
	\centering
	\includegraphics[width=1\textwidth]{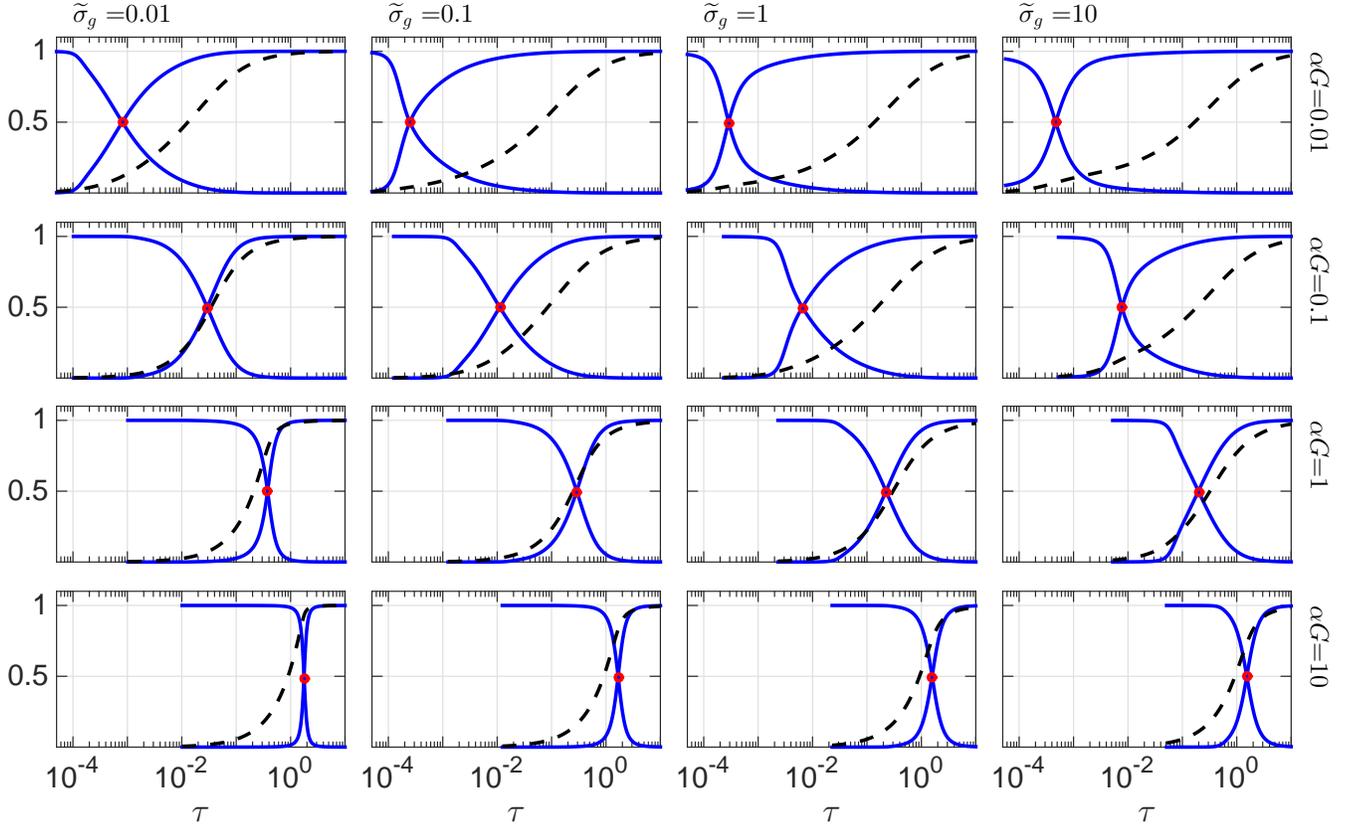}
	\caption{H{\tiny I} and \hh density profiles as functions of dust optical depth in a log-linear display.
	}
	\label{fig: profiles_tau}
\end{figure*}
\begin{figure*}
	\centering
	\includegraphics[width=0.7\textwidth]{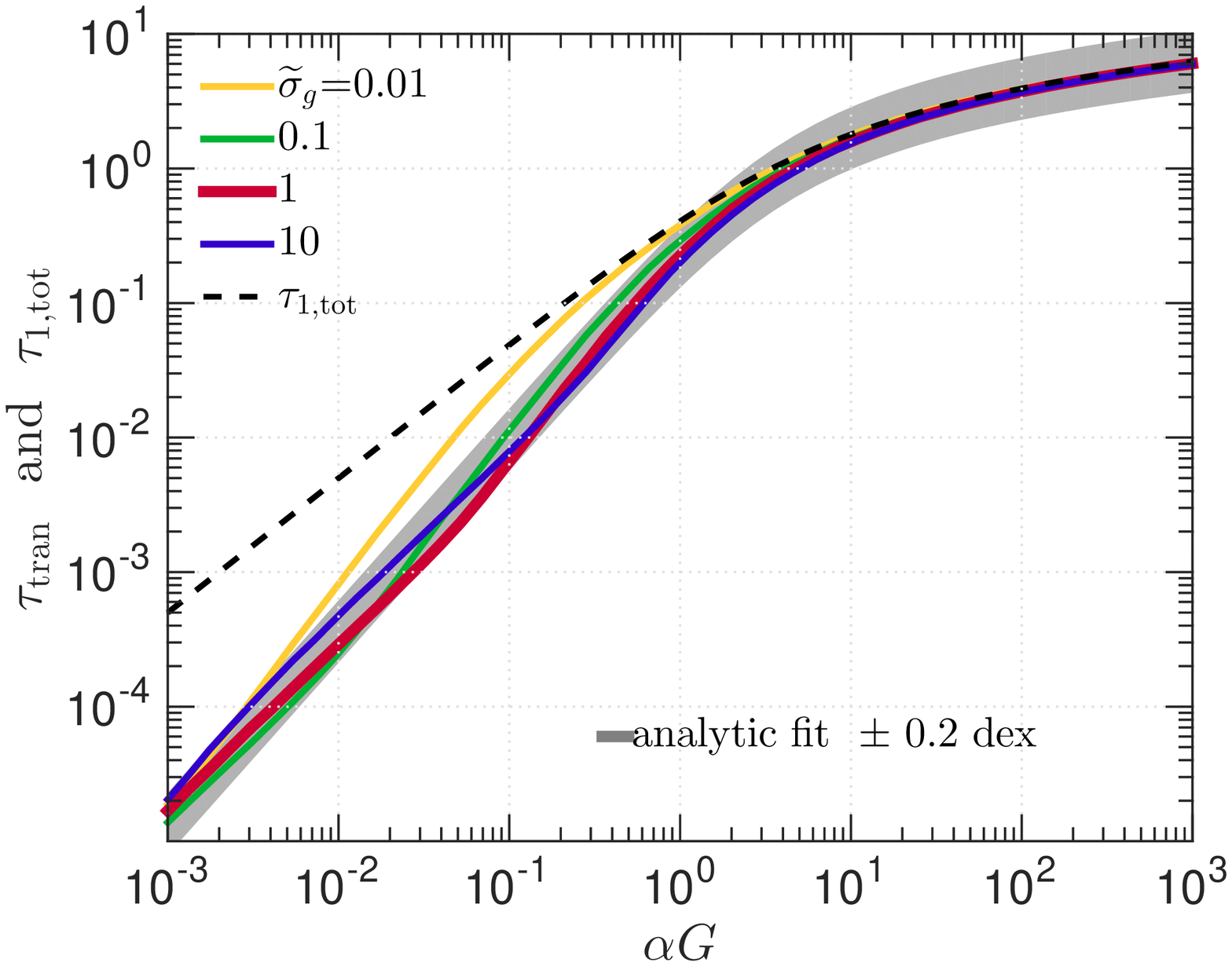}
	\caption{The optical depths at the H{\tiny I}-to-H$_2$ transition points $\tau_{\rm tran}$ (the red dots in Fig.~\ref{fig: profiles_tau_log}), as functions of the \aG parameter, for $\st=0.01$, 0.1, 1, and 10 (solid colored curves). 
For $\st \gtrsim 0.1$, $\tau_{\rm tran}$ depend only weakly on \st and is determined mainly by the \aG parameter.
The grey strip is our fitting function given by Equation (\ref{eq: tau_g_tran approx}) with $\beta=0.7$, with a width indicating $\pm 0.2$ dex variations from the formula.
	}
	\label{fig: tau1_transition}
\end{figure*} 

Starting with Equation (\ref{eq: profiles2}), we use our analytic procedure, to generate \ho and \hh density profiles for a wide range of $\aG$ and $\sg$.
We are particularly interested in the metallicity behavior so we consider large variations in the dust cross-section.
We present results for $4\times 4$ combinations of $\aG = 0.01, 0.1, 1$, and 10, and $\st = 0.01, 0.1, 1$, and 10, where \aG and \st are treated as independent variables.
We then use the profiles to derive an analytic scaling relation for the \tra transition points.

\subsection{\ho and \hh densities versus gas column}
\label{sub: Profiles vs N}

In Fig.~\ref{fig: profiles_N} we plot (blue solid curves) the atomic and molecular density fractions, $n_1/n$ and $2n_2/n$, as functions of the total gas column density $N$, for  
 all $4 \times 4$ combinations of $\aG$ and $\st$. 
 As shown by the curves in each panel, for sufficiently small $N$ the gas is entirely atomic, and for large $N$ the gas is molecular. 
We define the transition points as the depths at which $n_1=2n_2$.
These are indicated by the red dots in Fig.~\ref{fig: profiles_N}.

 For $N \rightarrow 0$ the ratio $n_1/n_2 \rightarrow \alpha/2 \equiv D_0/(2 R n)$, or equivalently $2n_2/n \rightarrow 4/\alpha$ since $n_1/n \simeq 1$ at the cloud boundaries.
 We stress that in each row in Fig.~\ref{fig: profiles_N} it is $\aG$ that is held constant, not $\alpha$.
For \aG constant, $\alpha$ must be altered with the dust cross-section as 
$(1+8.9\st)^{0.37}/\st$.
In Fig.~\ref{fig: profiles_N}, this is reflected in the increasing molecular fraction at the cloud boundary, as \st is increased.
For example, for $\aG=0.1$, and for $\st$ from 0.01 to 10, $\alpha$ decreases from $1.5 \times 10^{5}$ to $7.4 \times 10^{3}$.
The corresponding molecular fractions at the boundaries increase from $2n_2/n=2.7\times 10^{-5}$ to $5.3\times 10^{-3}$.

 
In each panel we also plot (dashed curves) the normalized accumulated \ho columns 
\begin{equation}
{\widetilde N_1} \ \equiv \ \frac{N_1(N)}{N_{1,{\rm tot}}} \ .
\end{equation}
Here $N_1(N)$ is obtained from step-3 of the procedure (\S \ref{sec: Profile constraction}), and $N_{\rm 1, tot}$ is the total \ho column (Equation (\ref{eq: N_1_tot})). 
The curves for ${\widetilde N_1}$ show how the \ho columns are built up with increasing cloud depth, 
finally reaching the maximal values when all of the photodissociating radiation has been absorbed.
As discussed by S14, in the strong-field limit ($\aG \gtrsim 1$) most of the \ho column density is built up in an outer fully photodissociated layer, prior to the conversion points (e.g.~see the lowest row in Fig.~\ref{fig: profiles_N}). These are the strong-field ``sharp" transitions. 
In contrast, in the weak-field limit ($\aG \lesssim 1$), the transitions are ``gradual" and most of the \ho column density is built up past the transition points where the gas is predominantly molecular (e.g.~see top row in Fig.~\ref{fig: profiles_N}).

In Fig.~\ref{fig: profiles_N}, for any $\st$, the transition points occur at greater cloud depths as \aG is increased, i.e.~as the free-space dissociation rate is increased relative to the H$_2$-formation rate. 
When $\aG$ is small, the transition points are controlled by H$_2$-self shielding.
For large $\aG$, the transition points are limited by \hd in addition to self-shielding.
For example, for fixed $\st=1$, the transition point occurs at gas columns $N_{\rm tran}=1.5\times10^{17}$, $3.5 \times 10^{18}$, $1.2\times 10^{20}$ , and $8.3\times10^{20}$~cm$^{-2}$, for $\aG=0.01$, 0.1, 1, and 10 (top to bottom in Fig.~\ref{fig: profiles_N}).

For fixed $\aG$, the transition points occur at greater gas columns as \st is reduced.
In the strong field limit ($\aG \gtrsim 1$), this is simply because the gas column required to achieve a large ($\gtrsim 1$) \hd optical depth grows as the dust absorption cross-section is reduced.
Thus, for example, for $\aG=10$, the transition column increases from $\sim 10^{20}$ to $10^{23}$~cm$^{-2}$, as \st is reduced from 10 to 0.01 (see bottom row of Fig.~\ref{fig: profiles_N}).

For small \aG the behavior is more complicated.
Dust absorption plays no role, but a dependence on \st remains.
For example, for $\aG=0.01$, the transitions occur at gas columns equal to $2.5\times 10^{16}, 1.5\times10^{17}, 1.3\times10^{18}$, and $4.3\times10^{19}$, as \st is reduced from 10 to 0.01 (see top row of Fig.~\ref{fig: profiles_N}).
This is because we have assumed that the grain surface H$_2$ formation rate coefficient, $R$, is proportional to $\st$ (see Equations (\ref{eq: R_H2}) and (\ref{eq: aG param})), with both quantities scaling in the same way with the dust-to-gas ratio.
Thus, for small \aG the transition points occur at larger gas columns as \st is reduced due to the associated reduction in the H$_2$ formation efficiency.
The behavior is non-linear however, because of the dependence on the H$_2$ self-shielding function.

Our analytically generated profiles in Fig.~\ref{fig: profiles_N}, and the profiles computed using the full multi-line radiative transfer in S14 (see their Fig.~7) are in excellent agreement.

\subsection{\ho and \hh densities versus dust opacity}
\label{sub: profiles vs tau}
The overall behavior becomes simpler if the density profiles are plotted as functions of the dust optical depth $\tau\equiv\sg N$, instead of the gas column $N$.
A very useful scaling relation for the transition point is also revealed by making this transformation.
In Fig.~\ref{fig: profiles_tau_log} we show $n_1/n$, $2n_2/n$ and $\widetilde{N}_1$ as functions of $\tau$ as log-log plots. 
In Fig.~\ref{fig: profiles_tau} we display the profiles as log-linear plots.
The transition points are again indicated by the red dots.

The dust optical depth at the transition is
\begin{equation}
\tau_{\rm tran} \ \equiv \ \sg (N_{\rm 1, tran}+2N_{\rm 2, tran}) \ ,
\end{equation}
where $N_{\rm 1, tran}$ and $N_{\rm 2, tran}$ are the integrated atomic and molecular columns up to the transition point (where $n_1=2n_2$).
As expected, for $\aG \gtrsim 1$ \hd absorption becomes significant and $\tau_{\rm tran} \gtrsim 1$.
For $\aG \ll 1$, self-shielding dominates and $\tau_{\rm tran} \ll 1$.
Remarkably, as can be seen by inspecting Fig.~\ref{fig: profiles_tau_log} or \ref{fig: profiles_tau}, for any $\aG$, $\tau_{\rm tran}$ is quite insensitive to $\st$, especially in the astrophysically important range 0.1 to 10.
Indeed to an excellent approximation, the profiles structures depend {\it only} on $\aG$. 
This is indicated by the almost identical profiles in each row of Fig.~\ref{fig: profiles_tau} in the linear-log displays that suppress the differences in the free-space \ho/\hh ratios near the boundaries.

This behavior is not surprising for large $\aG$, for which the transitions all occur at $\tau_{\rm tran} \sim 1$.
However, for small $\aG$ this result is less immediate, 
and depends on the complex interplay between H$_2$ self-shielding and the connection between the H$_2$ formation rate coefficient and the dust cross section, and on the definition of the effective dissociation bandwidth.

The precise locations of the transition points do also depend on the assumed absorption-line Doppler parameter when $\aG \ll 1$ and $\st \gtrsim 1$ since then self-shielding at the transition point occurs near the Doppler cores.
In Fig.~\ref{fig: profiles_tau_log}, the horizontal bars indicate the variation in the location of the transition points for $b_5=1$ to 4 (from left-to-right), a factor of two variation in our assumed Doppler parameter $b_5=2$.
As \st becomes small (still at $\aG \ll 1$), the self-shielding at the transition points occurs within the damping wings, and the \tra transition is then independent of $b$.

\subsection{Analytic expression for the transition point}
\label{sub: transition points}

That the transition optical depth $\tau_{\rm tran}$ depends mainly on the single parameter $\aG$ is a primary result of this paper, and we now present an analysis 
for this behavior.


In Fig.~\ref{fig: tau1_transition} we plot $\tau_{\rm tran}$ versus $\aG$ 
ranging from 10$^{-3}$ to 10$^3$, for the four values of $\st$. 
We computed these curves using our procedure for generating the depth-dependent profiles for any $\aG$, and then extracted the transition points. 
For comparison, we also plot (dashed curve) the total \hd optical depth $\tau_{\rm 1, tot}$ as given by Equation (\ref{eq: tau_1_tot}) (see also Fig.~\ref{fig: N1_tot_tau1_tot}).

For large $\aG$, the curves for $\tau_{\rm tran}$ all converge to $\tau_{\rm 1, tot}$ as expected in this limit for which the \tra transitions are sharp.
Thus for large $\aG$
\begin{equation}
\label{eq: tau_tran_large_aG}
\tau_{\rm tran} \ \simeq \ \tau_{\rm 1, tot} \ = \ \ln \Big[\frac{\aG}{2} + 1 \Big] \ \simeq \ \ln \Big[\frac{\aG}{2} \Big] 
\end{equation} 
and independent of $\st$.

For small $\aG$, $\tau_{\rm tran} \ll \tau_{\rm 1, tot}$, since the transitions are controlled by H$_2$-self shielding and most of the \ho columns are built up past the transition points.
Fig.~\ref{fig: tau1_transition} shows that $\tau_{\rm tran}$ remains insensitive to \st even as \aG becomes small.
For \st between 0.1 and 10, the transition optical depths do not deviate by more than a factor 2 for any \aG down to 10$^{-3}$.
For smaller \st the transition optical depths increase somewhat as indicated by the curve for $\st = 0.01$.

This behavior may be understood analytically as follows.
When \aG is small, $\tau \ll 1$, and dust absorption may be neglected 
at the transition point.
The \ho and \hh column densities at the transition are then given by
Equations (\ref{eq: n_1/n_2 alpha}) and (\ref{eq: N_1(N_2) no dust}), evaluated for $n_1=2n_2$, giving the pair of equations
\begin{equation}
\label{eq: small aG N2 tran}
\frac{n_1}{n_2} \ = \ 2 \ = \ \frac{1}{2} \alpha \ f(N_{2,{\rm tran}}) \ ,
\end{equation}
and
\begin{equation}
\label{eq: small aG N1 tran}
N_{1,{\rm tran}} \ = \ \frac{\alpha}{2} \ \frac{W_d(N_{2,{\rm tran}})}{\sigma_d} \ .
\end{equation}
To obtain simple solutions for the two unknowns $N_{\rm 1, tran}$ and $N_{\rm 2, tran}$, we approximate the H$_2$ shielding function as a power-law
 $f \propto N_2^{-\beta}$, so that $W_d \propto N_2^{1-\beta}$ (see Equation (\ref{eq: f_shield})).
 It follows that
 \begin{equation}
   N_{\rm 1, tran} \ \propto \ N_{2,{\rm tran}} \ \propto \ \alpha^{1/\beta} \ .
 \end{equation}
 The optical depth at the transition point then obeys
 \begin{equation}
 \label{eq: tau_tran small aG, in alpha}
 \tau_{\rm tran} \ \propto \ \st \ \alpha^{1/\beta} \ .
 \end{equation}
If expressed in terms of $\alpha$ alone $\tau_{\rm tran}$ is proportional to $\st$.

 To express Equation (\ref{eq: tau_tran small aG, in alpha}) in terms of $\aG$, we multiply and divide $\alpha$ by $G \propto \st (1+8.9 \st)^{-0.37}$.
 This gives
\begin{equation}
\label{eq: tau_g_tran small aG}
\tau_{\rm tran} \ \propto \ \st^{(\beta-1)/\beta} \ (1+8.9\st)^{0.37/\beta} \ (\aG)^{1/\beta} \ .
\end{equation} 
Thus, when considering $\tau_{\rm tran}$ in terms of \aG rather then $\alpha$, two important limits arise
\begin{equation}
\label{eq: tau_tran limits}
\tau_{\rm tran} \propto (\aG)^{1/\beta} \times \begin{dcases*}
\widetilde{\sigma}_g^{(\beta-0.63)/\beta} & for $ \st \gg 0.1$\\
\widetilde{\sigma}_g^{(\beta-1)/\beta} & for $ \st \ll 0.1 \ \ \ .$
\end{dcases*}
\end{equation}
As discussed in \S\ref{sub: fitting for f}, for $N_2$ in the range $10^{14} - 10^{21}$~cm$^{-2}$,  $f_{\rm shield} \propto N_2^{-\beta}$ with $\beta \simeq   0.7$ is a good approximation.
The dependence of $\tau_{\rm tran}$ on \st is then very weak, especially for $\st \gtrsim 0.1$, for which $\tau_{\rm tran} \propto \st^{0.10}$.
Furthermore, in this approximation $\tau_{\rm tran}$ varies as a power-law $(\aG)^{1/\beta}$ as seen in Fig.~\ref{fig: tau1_transition}.


An excellent fitting formula for $\tau_{\rm tran}$ as a function of \aG only is therefore
\begin{equation}
\label{eq: tau_g_tran approx}
\tau_{\rm tran} \ = \ \beta  \ \ln\Big[ \Big(\frac{\aG}{2}\Big)^{1/\beta}+1\Big] \ .
\end{equation} 
For $\aG \gg 1$, this gives $\tau_{\rm tran} \rightarrow \ln[\aG/2]$ as it should (Equation (\ref{eq: tau_tran_large_aG})).
For $\aG \ll 1$ the power-law behavior $\tau_{\rm tran} \propto (\aG)^{1/\beta}$ is recovered (Equation (\ref{eq: tau_g_tran small aG})), with prefactor $\beta$ of order unity.
The grey strip in Fig.~\ref{fig: tau1_transition} shows $\tau_{\rm tran}$ as given by Equation (\ref{eq: tau_g_tran approx}) with $\beta =0.7$, with a width indicating $\pm 0.2$ dex variations from the formula.
This choice of $\beta$ provides best accuracy for $\st \geq 0.1$.
The maximal absolute deviations over the whole $\aG$ range are 0.16, 0.21 and 0.20 dex, and the medians (in log space) are 0.03, 0.02 and 0.04 dex, for $\st = 0.1$, 1, and 10, respectively.
For $\st =0.01$, the median and maximum absolute deviations are 0.15 dex and 0.49 dex.
For factor two variations in the Doppler parameter $\beta$ varies by less than 5 \%.

We stress that our universal scaling relation for $\tau_{\rm tran}$ depends on the inclusion of the H$_2$-dust absorption term in the definition of $\aG$, i.e.~the $(1+8.9\st)^{-0.37}$ factor.
Without this metallicity-dependent term built into the definition of $\aG$, the transition point would depend on $\aG$ and $\sg$ independently.

As examples for how our formula may be used, we compute the gas column density for the transition point for an incident radiation field with $I_{\rm UV}=1$, and cloud gas density $n=10^3$~cm$^{-3}$, assuming a standard gas-to-dust ratio, i.e.~$\sg=1.9 \times 10^{-21}$~cm$^2$, and $R=3 \times 10^{-17}$~cm$^3$~s$^{-1}$.
For these conditions $\st=1$, 
and $\aG = 5.7 \times 10^{-2}$.
Plugging into Equation (\ref{eq: tau_g_tran approx}) with $\beta=0.7$ gives $\tau_{\rm tran} = 4.3 \times 10^{-3}$, or $N_{\rm tran} = 2.3 \times 10^{18}$~cm$^{-2}$.
This is only 15 \% of the total \ho column for this value of $\aG$, which is $1.5 \times 10^{19}$~cm$^{-2}$.
Reducing the metallicity by a factor of 10, and with the assumption that this implies a linear decrease in the dust cross section, this gives $\st=0.1$ 
and $\aG=0.14$, for which $\tau_{\rm tran} = 1.5 \times 10^{-2}$ and $N_{\rm tran} = 8.2 \times 10^{19}$~cm$^{-2}$.

For $\aG = 1$, the critical value between the weak- and strong-field limits, $\tau_{\rm tran} = 0.22$, and is approximately half the value of $\tau_{\rm 1, tot}$.

\subsection{Attenuation factor}
\label{sub: Attenuation factors}
\begin{figure*}[t]
 	\centering
	\includegraphics[width=1\textwidth]{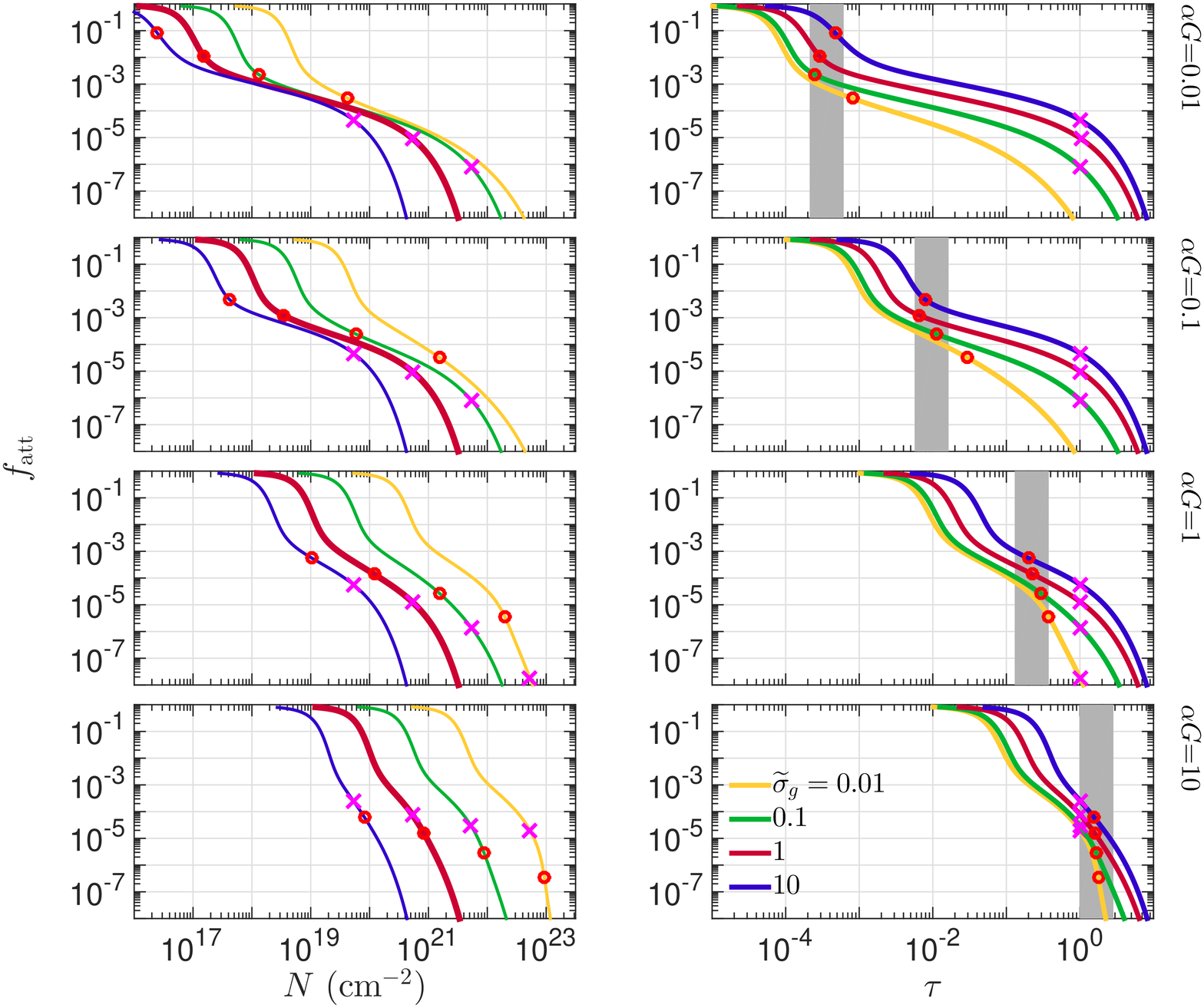}
	\caption{The attenuation factor $f_{\rm att}$ as a function of the gas column $N$ (left) and the optical depth $\tau$ (right), for $\aG =0.01, 0.1, 1$ and 10 (top to bottom panels), and for $\st = 0.01, 0.1, 1$ and 10 (color coded). The red circles indicate the H{\tiny I}-to-H$_2$ transition points. The magenta crosses indicate points of $\tau=1$.
	}
	\label{fig: shielding_factor}
\end{figure*}
Given solutions for the depth dependent atomic and molecular density ratios $n_1/n_2$, we can compute the corresponding depth-dependent attenuation factors for the H$_2$ photodissociation rate.
These are given by
\begin{equation}
f_{\rm att}(N; \aG ,\st) \ = \ \frac{2}{\alpha} \ \frac{n_1}{n_2} (N) \  ,
\end{equation} where as discussed above, any pair $(\aG, \st)$ gives $n_1/n_2$ versus $N$, and determines $\alpha$.
In Fig.~\ref{fig: shielding_factor} we plot $f_{\rm att}$ as a function of the total gas column $N$ (left panels) and of the dust opacity $\tau \equiv \sg N$ (right columns).
Each panel is for a fixed $\aG$, and shows four curves for the different \st values.
The transition points are marked by the red dots. The crosses indicate the points where the dust opacity  $\tau=1$.

The curves show the combined effects of H$_2$-self shielding and dust absorption in reducing the photodissociation rate (recall $f_{\rm att} \equiv f_{\rm shield} \mathrm{e}^{-\tau}$).
For small \aG (e.g.~upper panel) the curves for $f_{\rm att}(N)$ resemble $f_{\rm shield}(N_2)$
(see Fig.~\ref{fig: Wg}) for much of the range, because the \tra transition points
occur at small gas columns, well before any dust opacity is built up. The sudden
onset of self-shielding and drop in the photodissociation rate is followed
by a more gradual decline as the H$_2$ photoabsorption occurs in the line
damping wings. Finally, at sufficiently large columns $f_{\rm att}$ drops exponentially due to
the H$_2$-dust cutoff, or due to the complete overlap of the absorption lines.
For large \aG (e.g.~lower panel) the exponential \hd cutoff occurs
soon after the onset of self-shielding, and the damping wing portions
are suppressed.  For large $\aG$, most of the LW-band radiation is
absorbed by $\hd$. However, self-shielding is still important in reducing
the photodissociation rate significantly and enhancing the H$_2$ fraction
within the large atomic layer.

In the right-hand panels of Fig.~\ref{fig: shielding_factor} the vertical grey strips show the
predicted positions of the transition points as given by our
Equation (36), with a width $\pm 0.2$ dex (as for Fig.~\ref{fig: tau1_transition}).
For \st between 0.1 and 10, the computed transition points
fall within the strips.  For $\st=0.01$ the transition points
occur at slightly larger dust optical depths, as we have
already seen in Fig.~\ref{fig: tau1_transition}.

\section{H{\tiny I}-to-H$_2$ star-formation threshold}
\label{sec: star-formation}

\begin{figure}[t]
	\centering
	\includegraphics[width=0.5\textwidth]{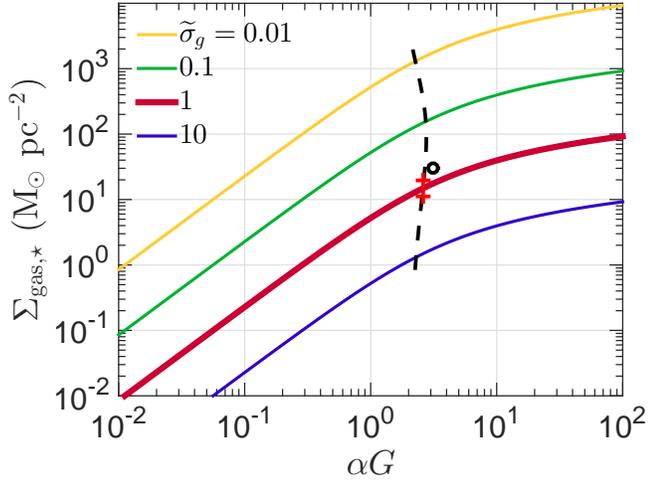}
	\caption{Star-formation thresholds as functions of $\aG$, for $\st=0.01$, 0.1, 1, and 10.
	The dashed curve is the locus $\aG = (\aG)_{\rm CNM}$ (see text).
	The red crosses are the thresholds predicted by S14 for beamed (upper) and isotropic (lower) irradiation. The black circle is the threshold derived by \citet{McKee2010}.  
	}
	\label{fig: thresholds}
\end{figure}

We can use our Equation (\ref{eq: tau_g_tran approx}) to define a star-formation threshold with the empirically based assumption that conversion to H$_2$ is required for star-formation, the \ho remaining sterile.
Following S14 we define the threshold gas mass surface density, $\Sigma_{\rm gas ,\star}$ (M$_{\odot}$~pc$^{-2}$), as the surface density required for half of the mass to be molecular. 
For our idealized plane-parallel clouds, and for two-sided illumination, the threshold hydrogen gas column $N_{\rm gas, \star} \equiv 4 N_{\rm tran} = 4 \sg \tau_{\rm tran}$.
Here we are assuming that once the gas becomes predominantly molecular stars can form, even if residual \ho exists in the molecular zone as occurs in the weak field limit.
Converting to a mass surface density, including the contribution of helium to the mass, gives
\begin{equation}
\label{eq: SF threshold}
\Sigma_{\rm gas, \star} \ = \ \frac{16.5}{\st} \ \ln\Big[ \Big(\frac{\aG}{2} \Big)^{1.43} + 1 \Big] \ \ {\rm M_{\odot} \ pc^{-2}} \  ,
\end{equation} where we set $\beta=0.7$ in Equation (\ref{eq: tau_g_tran approx}).
In Fig.~\ref{fig: thresholds} we plot the star-formation thresholds as functions of $\aG$, for $\st$ from 0.01 to 10.
We recall that the dust absorption cross-section may be related to the metallicity, (Equation (\ref{eq: sg Z})) with $\sg=1.9 \times 10^{-21} \phi_g Z'$~cm$^2$, or $\st=\phi_g Z'$, where $\phi_g$ is a factor of order unity.

For large \aG (strong field limit), the \ho layers become extended but are limited by \hd absorption.
For example for $\st=1$ (or $Z' \approx 1$), $\Sigma_{\rm gas, \star}$ increases from 5 to 92 M$_{\odot}$~pc$^{-2}$ for \aG from 1 to 10$^2$.
For low $\aG$ (weak field limit) the outer \ho layers become very narrow and $\Sigma_{\rm gas, \star}$ drops sharply. For $\st=1$, $\Sigma_{\rm gas, \star}$ decreases from 5 to $8 \times 10^{-2}$ M$_{\odot}$~pc$^{-2}$ for $\aG=1$ to 10$^{-2}$.
Because the dust optical depth at the \tra conversion point is insensitive to $\sg$, the threshold surface-density varies with metallicity as $1/Z'$ for any $\aG$.

If the \ho gas is assumed to be cold (CNM) at multiphased conditions (\citealt{Krumholz2008}; S14) then $\aG=(\aG)_{\rm CNM}$ in Equation (\ref{eq: SF threshold}), where as given by S14\footnote{In this expression we have adjusted Eq.~(59) of S14 for $(\aG)_{\rm CNM}$ by using our adopted formula Eq.~(\ref{eq: Wg_tot}) for $W_{g,{\rm tot}}$, instead of Eq.~(\ref{eq: Wg_tot_S14})}
\begin{equation}
(\aG)_{\rm CNM} \ \equiv \ 2.6 \Big(\frac{1+3.1 Z'^{0.365}}{4.1} \Big) \Big( \frac{9.9}{1+8.9 \phi_g Z'} \Big)^{0.37}\ .
\end{equation}
The dashed curve in Fig.~\ref{fig: thresholds} shows the locus $\aG=(\aG)_{\rm CNM}$ for $Z'$ from 10 to 0.01 (assuming $\phi_g=1$).
For $\st=1$ ($Z'\approx1$), and for $\aG=(\aG)_{\rm CNM}$, the threshold surface density is 15 M$_{\odot}$~pc$^{-2}$.

S14 gave expressions for the thresholds $\Sigma_{\rm gas, \star}$ but with $\tau_{\rm tran}$ replaced by $\tau_{\rm 1, tot}$, with the added assumption (valid for $\aG \gtrsim 1$) that the transitions are sharp. The upper and lower red crosses in Fig.~\ref{fig: thresholds} are the predicted S14 thresholds, equal to 11 and 20~M$_{\odot}$~pc$^{-2}$, for $\aG=(\aG)_{\rm CNM}$ and $\st=1$, for beamed and isotropic irradiation respectively. 
The circle shows the \citet{McKee2010} result of 30~M$_{\odot}$~pc$^{-2}$, for their assumed $(\aG)_{\rm CNM}=3.1$ for solar metallicity, and assuming spherical geometry\footnote{S14 present (\S 4) a detailed comparison between the formulae and results for slab geometry versus the spheres assumed by \citet{Krumholz2009, McKee2010}.}.

In general, the assumption that $\aG=(\aG)_{\rm CNM}$ may be too restrictive, e.g.~as shown for the \tra transitions in the Perseus molecular cloud \citep{Bialy2015b}.
Our Equation (\ref{eq: SF threshold}) provides estimates for the thresholds for arbitrary \aG and metallicity.

\section{Summary}
\label{sec: discussion}
In this paper we have extended the analytic theory presented by \citet[][, S14]{Sternberg2014} for the \tra transitions in interstellar photon dominated regions (PDRs), to develop a simple procedure for the construction of steady-state \tra density profiles for FUV irradiated clouds.
Following S14, we first demonstrate that for uniform density gas, and for a steady irradiation fluxes, the atomic to molecular gas fractions as functions of cloud depth, in one-dimensional slabs, depend on two basic parameters. First is the dimensionless ``$\aG$ parameter" (\citealt{Sternberg1988}; S14) that determines the dust optical depth associated with the total photodissociated \ho column.
Second is $\sg$, the dust-grain absorption cross-section per hydrogen nucleus in the 1100 to 912 $\AA \ $ Lyman-Werner  photodissociation band. 
The \aG parameter is proportional to the ratio of the FUV field intensity to the gas density, or alternatively to the ratio of the H$_2$ dissociation rate to the molecular formation rate (see Equations (\ref{eq: aG}) and (\ref{eq: aG param with R})).
The dust-grain absorption cross-section also enters into the definition of $\aG$ because some of the LW band radiation may be absorbed by dust associated with just the H$_2$.

We then develop our analytic procedure for generating \tra density profiles.
It may be used for arbitrary input gas density, FUV field intensity, H$_2$ formation rate coefficient and dust-grain absorption cross-section.
The gas density and field intensity may vary widely in differing interstellar environments.
The formation rate coefficient and dust absorption cross-section will generally vary with the metallicity dependent dust-to-gas ratio.
Our simple 5-step procedure (\S \ref{sec: Profile constraction}) requires analytic forms for (a) the H$_2$-self shielding function, (b) the dust limited curve-of-growth for multiline dissociation bandwidth, and (c) the total dust-limited bandwidths.
We develop the required analytic expressions for (b) and (c) (Equations (\ref{eq: fit}) and (\ref{eq: Wg_tot})) based on the detailed radiative transfer results presented by S14.
For (a) we recommend the \citet{Draine1996} formula (Equation (\ref{eq: f_shield DB96})).

We use our procedure to generate a $4\times 4$ set (Fig.~\ref{fig: profiles_N})
of \tra transition profiles, showing the fractions as functions of the total gas columns, from large (strong-field) to small (weak-field) $\aG$, and for dust cross-sections ranging from super-solar to very  low dust-to-gas mass ratios. We then show that if the profiles are displayed (Figs.~\ref{fig: profiles_tau_log}-\ref{fig: profiles_tau}) as functions of the dust optical depth (or visual extinction) the dependence on \sg becomes very weak, and just $\aG$ remains as the single controlling parameter, including also for the depth at which the \tra transition occurs.
This simple scaling behavior is valid from the strong-field limit where \hd opacity dominates at the transition point, to the weak-field limit where the transition is controlled by self-shielding only.

We use our results to derive a simple universal fitting formula for the dust optical depth at which the \tra transition occurs, as a function of the single dimensionless parameter $\aG$ (Equation (\ref{eq: tau_g_tran approx}) and Fig.~\ref{fig: tau1_transition}).
We use our formula for the transition depth to develop an expression for the threshold gas mass surface density required for star-formation, under the assumption that individual clouds become star-forming when $\sim 50$~\% of the gas mass is molecular (Equation (\ref{eq: SF threshold}) and Fig.~\ref{fig: thresholds}).

Our analytic procedure and formulae for the \tra transition points and star-formation thresholds will be useful for interpreting observations of interstellar H{\small I}/\hh interfaces on small and global galactic scales, and for incorporation into hydrodynamics simulations.

 \begin{acknowledgements}
We thank Frank Le Petit and Evelyne Roueff for in-depth discussions about this work, and the referee for helpful comments that improved our presentation.
We thank Chris McKee for comments, and Henrik Beuther and Thomas Henning for stimulating conversations.
We thank Or Shrim for his assistance in generating the \ho and \hh density profiles.
S.B.~acknowledges support from the Raymond and Beverly Sackler Tel Aviv University -- Harvard/ITC Astronomy Program.
This work was supported 
by the PBC Israel Science Foundation I-CORE Program grant 1829/12.
\end{acknowledgements}

\end{document}